\documentclass[nofootinbib,prd,superscriptaddress,reprint]{revtex4-1}

\usepackage{amsmath,amssymb}
\usepackage{mathtools,color}
\usepackage[percent]{overpic}

\newcommand\gmunu{g_{\mu\nu}}
\newcommand\bmunu{\gamma_{\mu\nu}}
\newcommand\hmunu{h_{\mu\nu}}
\newcommand\Bmunu{B_{\mu\nu}}
\newcommand\PP{\mathbf{P}}
\newcommand\JJ{\mathbf{J}}
\newcommand\pp{\mathbf{p}}
\newcommand\ppbar{\mathbf{\bar{p}}}
\newcommand\ahat{\mathbf{\hat{a}}}
\newcommand\xhat{\mathbf{\hat{x}}}
\newcommand\yhat{\mathbf{\hat{y}}}
\newcommand\zhat{\mathbf{\hat{z}}}
\newcommand\ehat{\mathbf{\hat{e}}}

\newcommand\pwket{|\pp \ \lambda\rangle}
\newcommand\pwketbar{|\bar{\pp} \ \bar{\lambda}\rangle}

\begin{document}
\title{Quantum Emulation of Gravitational Waves}
\author{Ivan \surname{Fernandez-Corbaton}\footnote{ivan.fernandez-corbaton@kit.edu; Corresponding author.}}
\email{ivan.fernandez-corbaton@kit.edu}\thanks{Corresponding author.}

\affiliation{Department of Physics \& Astronomy, Macquarie University, Australia}
\affiliation{ARC Center for Engineered Quantum Systems, Macquarie University, North Ryde, New South Wales 2109, Australia}
\affiliation{Institute of Nanotechnology, Karlsruhe Institute of Technology, 76021 Karlsruhe, Germany}
\author{Mauro Cirio}
\affiliation{Department of Physics \& Astronomy, Macquarie University, Australia}
\affiliation{ARC Center for Engineered Quantum Systems, Macquarie University, North Ryde, New South Wales 2109, Australia}
\author{Alexander B\"use}
\affiliation{Department of Physics \& Astronomy, Macquarie University, Australia}
\affiliation{ARC Center for Engineered Quantum Systems, Macquarie University, North Ryde, New South Wales 2109, Australia}
\author{Lucas Lamata}
\affiliation{Department of Physical Chemistry, University of the Basque Country UPV/EHU, Apartado 644, E-48080 Bilbao, Spain}
\author{Enrique Solano}
\affiliation{Department of Physical Chemistry, University of the Basque Country UPV/EHU, Apartado 644, E-48080 Bilbao, Spain}
\affiliation{IKERBASQUE, Basque Foundation for Science, Maria Diaz de Haro 3, 48013 Bilbao, Spain.}
\author{Gabriel Molina-Terriza}
\affiliation{Department of Physics \& Astronomy, Macquarie University, Australia}
\affiliation{ARC Center for Engineered Quantum Systems, Macquarie University, North Ryde, New South Wales 2109, Australia}

\begin{abstract}
Gravitational waves, as predicted by Einstein's general relativity theory, appear as ripples in the fabric of spacetime traveling at the speed of light. We prove that the propagation of small amplitude gravitational waves in a curved spacetime is equivalent to the propagation of a subspace of electromagnetic states. We use this result to propose the use of entangled photons to emulate the evolution of gravitational waves in curved spacetimes by means of experimental electromagnetic setups featuring metamaterials.

\end{abstract}
\maketitle
\section*{Introduction}
Gravitational waves are commonly described as ripples in the fabric of spacetime that travel at the speed of light. According to general relativity, their direct detection on Earth will only be possible for waves created when spacetime is stirred by the movement of very massive astrophysical objects, like black holes and neutron stars \cite[\S 36.3-6]{Misner1973}. We have indirect evidence of the existence of gravitational waves \cite{Weisberg1981,Taylor1982}, but they have not yet been detected directly. The considerable efforts placed in developing and building detectors \cite{Barish1999} should allow the direct observation of these elusive objects in the near future \cite{Ju2000}. The analysis of gravitational radiation should provide a completely new source of information about our universe. For example, while we will never have access to electromagnetic information older than the Cosmic Microwave Background, we may one day be able to study gravitational waves produced at earlier epochs, much closer to the  Big Bang \cite{Thorne1997}.

From the theoretical side, gravitational waves are modeled as propagating perturbations of the spacetime metric. A wide class of approximate gravitational wave solutions can be studied by means of Isaacson's high frequency limit \cite{Isaacson1968,Isaacson1968b}. This limit is analogous to the geometric-optics approximation of Maxwell's equations. The analogy between gravitation and electromagnetism goes well beyond this particular mathematical approach. Several different analogies are collected under the name {\em gravitoelectromagnetism}. We refer the reader to \cite[II.3]{Iorio2007} for a concise review with numerous references. The analogies range from the similarity between Coulomb's law of electricity and Newton's law of gravitation, through the use of electric like and magnetic like vectors to approximately describe general relativity, to the non-perturbative covariant formulation of Maxwell-like forms of gravitational tensors and their dynamical equations \cite{Maartens1998}.

In this article, we develop an equivalence between the propagation of gravitational waves in a curved spacetime background and the propagation of a restricted set of electromagnetic tensor waves in the same background. The factorization of a circularly polarized gravitational wave as the tensor product of two circularly polarized electromagnetic waves, which is not possible for linear polarizations, is central to this equivalence. Its physical meaning becomes clear when treating polarization by means of the helicity operator and its eigenstates. As far as we know, this equivalence has not been reported before.

After establishing the equivalence, we use it to propose a path for the emulation of the propagation of gravitational waves in a curved spacetime background by means of electromagnetic setups in a laboratory. To achieve the proposed equivalence for all regimes, it is necessary to use a genuine quantum system, i.e. entangled photons. Therefore, we propose the simulation of a purely classical system with a genuinely quantum setup. Conformal invariance is used to shrink the astronomical sizes down to laboratory scales, and a suitable metamaterial media to mimic spacetime curvature. Up to our knowledge, this is a new emulation approach, but we point out that there has been an intensive research on quantum simulations of gravitational phenomena like black holes and gravitational waves in Bose-Einstein condensates \cite{Garay2000,Lahav2010,Bravo2014}. In parallel, several proposals for detecting gravitational waves using cold atoms have been made \cite{Graham2013,Sabin2014}.

\section*{Physics of gravitational waves}

Let us first of all outline the gravitational wave physics that we want to capture. We adopt the conventions of Einstein's summation over repeated indexes, a flat metric $\eta_{\mu\nu} =  \rm diag (1,-1,-1,-1)$, and bold symbols to denote spatial three-vectors.

The first step for studying the propagation of gravitational waves is to decompose the total metric tensor $\gmunu$ into a slowly varying background $\bmunu$ and a rapidly varying wave $\hmunu$ \cite{Isaacson1968},\cite[\S 35.13-14]{Misner1973}:
\begin{equation}
		\gmunu=\bmunu+\hmunu.
\end{equation}
The gravitational wave $\hmunu$ is assumed to be a small perturbation on top of the background $\bmunu$, and the wavelengths in $\hmunu$ are assumed to be much smaller than any inhomogeneity scale of the background. More explicitly, if $\lambda_h$ is the largest wavelength contained in $\hmunu$, and $L_\gamma$ the characteristic distance over which the background changes, we will assume throughout the paper that
\begin{equation}
	\label{eq:regime}
	\frac{\lambda_h}{L_\gamma}\ll 1,
\end{equation}
which, for a fixed $\lambda_h$, limits the maximum curvature of the background which is of order $L^{-2}_\gamma$.

Under these restrictions, the propagation of $\hmunu$ in empty space follows the wave equation \cite[Eq. 5.12]{Isaacson1968}:
\begin{equation}
		\label{eq:delta}
		\Delta_{\{\gamma\}} \hmunu=0,
\end{equation}
where $\Delta_{\{\gamma\}}$ is the d'Alembertian operator in the spacetime geometry determined by $\bmunu$. In flat spacetime it reduces to the common form of the wave operator $\Delta_{\{\eta\}}=\partial_t^2/c-\sum_{i=1}^{3}\partial_{x_i}^2$, where $c$ is the speed of light which we will set to 1 from now on. Equation (\ref{eq:delta}) already assumes the choice of a Lorenz and traceless gauge: $D_\nu h^{\mu\nu}=0$, $\bmunu h^{\mu\nu}=0$, where $D_\nu$ denotes the covariant derivative \cite[Eqs. 5.8, 5.9]{Isaacson1968}. Exhausting the last remaining gauge freedom, we set $h_{\mu0}=0$ and turn $\hmunu$ into a purely spatial symmetric tensor (the symmetry $h_{\mu\nu}=h_{\nu\mu}$ is a gauge independent property of any metric tensor). This collection of gauge choices is known as the transverse traceless gauge, and selects the gauge invariant part of the wave \cite[\S 35.4]{Misner1973}, thereby isolating its physical degrees of freedom. We will work in the transverse traceless gauge throughout the article. 

Exploiting the assumption that $\bmunu$ varies much slower than $\hmunu$, approximate solutions of Eq. (\ref{eq:delta}) are obtained in \cite{Isaacson1968} with the ansatz:
\begin{equation}
	\label{eq:ansatz}
		\hmunu(x_\alpha)=\Bmunu(x_\alpha)\exp(i\phi(x_\alpha)),
\end{equation}
where $\Bmunu$ changes slowly with the spacetime coordinates $x_\alpha$, and $\phi$ has a possibly large first derivative $p_\mu(x_\alpha)=\partial_{\mu}\phi(x_\alpha)$, which also changes slowly with $x_\alpha$, and no significant derivatives of higher orders. In this complex notation, the metric tensor is obtained from the real part of $\hmunu$. Substituting Eq. (\ref{eq:ansatz}) into Eq. (\ref{eq:delta}) leads to the important results that the $p_\mu$ are null vectors ($p_\mu(x_\alpha) p^{\mu}(x_\alpha)=0$), gravitational waves are transverse ($p_\mu(x_\alpha) B^{\mu\nu}(x_\alpha)=0$), and that their propagation can be understood as the parallel transport of both $p_\mu$ and $\Bmunu$ along the null geodesics $dx^\mu/dl=p^{\mu}$. Indexes are lowered and raised using $\bmunu$.

Identical results can be reached for the propagation of electromagnetic waves in curved spacetime \cite[Box 22.4]{Misner1973}, except that instead of the symmetric tensor $\hmunu$, they apply to the electromagnetic four-vector potential $A_\mu$. We will now show that the analogy with electromagnetism is exact when, instead of considering one instance of the electromagnetic field ($A_\mu$), one considers a subspace of the tensor products of two instances of it ($A_\mu A_\nu$). As we will show later, the states in this subspace can be physically realized by two-photon states.

\section*{Relationship between gravitational waves and electromagnetic waves in flat spacetime}

We start by considering solutions of Eq. (\ref{eq:delta}) in a flat background, i.e. $\bmunu=\eta_{\mu\nu}$. We will study the generalization to curved backgrounds in the next section. In a flat metric, plane waves constitute a complete basis of the space of propagating solutions. Plane waves are characterized by a null four-momentum and a constant polarization tensor. For each four-momentum $p_\mu$, there are only two independent polarization tensors that meet the constraints of the transverse traceless gauge. For $p_\mu=(\omega,0,0,\omega)$, the tensors that are typically chosen are \cite[Sec. 6]{Carroll1997}:
\begin{equation}
	\label{eq:sms}
		s=\xhat\otimes\xhat-\yhat\otimes\yhat, \text{ and }	m=\xhat\otimes\yhat+\yhat\otimes\xhat,
\end{equation}
where $\otimes$ denotes the tensor product. The tensors in Eq. (\ref{eq:sms}) are called the ``plus'' and ``cross'' linear polarizations. Another possible choice is the combinations $s\pm im$, which are referred to as circular polarizations. There is a key difference between these two sets of tensors: The linear polarization tensors $s$ and $m$ cannot be written as the tensor product of a vector with itself $\ahat\otimes\ahat$, but the $s\pm im$ combinations can
\begin{equation}
	\label{eq:fact0} 
		\begin{split}
			s+im=&\xhat\otimes\xhat-\yhat\otimes\yhat + i(\xhat\otimes\yhat+\yhat\otimes\xhat)=\\&(\xhat+ i\yhat)\otimes(\xhat+i\yhat),\\
			s-im=&\xhat\otimes\xhat-\yhat\otimes\yhat - i(\xhat\otimes\yhat+\yhat\otimes\xhat)=\\&(\xhat-i\yhat)\otimes(\xhat-i\yhat).
		\end{split}
\end{equation}
This factorization has notable implications. We can use Eq. (\ref{eq:fact0}) to factorize the complete expressions of the corresponding gravitational plane waves into a tensor product of two vector plane waves

\begin{eqnarray}
	\label{eq:splita}
	&&\left[(\xhat+ i\yhat)\otimes(\xhat+ i\yhat)\right]\exp(i\omega(z-t))=\\
	\nonumber
	&&\left[(\xhat+ i\yhat)\exp(i\kappa^+\omega(z-t))\right]\otimes\left[(\xhat+i\yhat)\exp(i(1-\kappa^+)\omega(z-t))\right],\\
	\label{eq:splitb}
	&&\left[(\xhat- i\yhat)\otimes(\xhat- i\yhat)\right]\exp(i\omega(z-t))=\\
	\nonumber
	&&\left[(\xhat- i\yhat)\exp(i\kappa^-\omega(z-t))\right]\otimes\left[(\xhat-i\yhat)\exp(i(1-\kappa^-)\omega(z-t))\right],
\end{eqnarray}
where $\kappa^{\pm}$ are scalars whose allowed values and significance will be discussed later. The vectorial waves have polarization vectors $(\xhat\pm i\yhat)$ which are transverse (orthogonal) to their four-momentum, which is proportional to that of the gravitational wave and hence also of null length. We recognize the expressions of the constituent vectorial waves in Eqs. (\ref{eq:splita}) and (\ref{eq:splitb}) as the spatial part of the four-vector potential of circularly polarized electromagnetic plane waves in the transverse gauge ($\nabla\cdot \mathbf{A}=0$). In this gauge, $\mathbf{A}$ collects the transverse degrees of freedom of the electromagnetic field. These physically relevant components are radiative, and are said to belong to the ``free field''. On the other hand, the non-radiative longitudinal degrees of freedom of the field can always be attached to the sources \cite[I.B.5]{Cohen1997},\cite[Chap. XXI,\S22]{Messiah1958}. In the transverse gauge the longitudinal components are contained in $A_0$, the time component of the vector potential. By setting $A_0=0$, the resulting electromagnetic gauge has the same physical significance as the  transverse traceless gauge in the gravitational case. This is seen by noting that with $A_0=0$, the transverse gauge is equivalent to the Lorenz gauge ($\partial_\mu A^{\mu}=0$), and that the zero trace and purely spatial conditions eliminate the non-radiative and longitudinal degrees of freedom of the gravitational field \cite[\S 35.4]{Misner1973}. 

Up to this point, we have just formally factorized two particular gravitational plane waves as tensor products of electromagnetic plane waves. We will now use a constructive procedure to show that there exists a formal equivalence between each gravitational plane wave and a set of physically valid entangled photon states.

Let us denote by $\pwket$ an electromagnetic plane wave with momentum $\pp$ and helicity $\lambda=\pm 1$. The helicity operator is the projection of the angular momentum operator vector in the direction of the linear momentum operator vector $\Lambda=\JJ\cdot\PP/|\PP|$ \cite[Chap. 8.4.1]{Tung1985}. Its eigenstates have the same polarization handedness in all their momentum components. Any valid electromagnetic tensor wave can be written as a linear superposition of states of the kind:
\begin{equation}
		\label{eq:boson}
		\pwket\otimes\pwketbar+\pwketbar\otimes\pwket,
\end{equation}
where the correct exchange symmetry for bosonic fields is enforced in an obvious way.

From the point of view of field theory, an object with zero mass and allowed helicity eigenvalues $\lambda=\pm 1$ is associated with the electromagnetic field, while an object with zero mass and allowed helicity eigenvalues $\lambda=\pm 2$ is associated with the gravitational field \cite[Chap. 2.5]{Weinberg1995}. Consequently, for electromagnetic states of the kind (\ref{eq:boson}) to be equivalent to gravitational waves, they must have their associated characteristics, i.e., zero mass and allowed helicity eigenvalues equal to $\pm 2$.

Let us first apply the mass squared operator to Eq. (\ref{eq:boson}):
\begin{equation}
	\label{eq:m}
	-\left(\prescript{}{2}{P}_0^2-\sum_{i=1}^3 \prescript{}{2}{P}_i^2\right)\left(\pwket\otimes\pwketbar+\pwketbar\otimes\pwket\right),
\end{equation}
where $\prescript{}{2}{P}_\mu$ is the generator of translations along direction $\mu$ in the tensor product space. Its expression as a function of $P_\mu$, the momentum operator for a single constitutive space, is: $\prescript{}{2}{P}_\mu=P_\mu\otimes I + I\otimes P_\mu$, where $I$ is the identity operator. Using that
\begin{equation}
	\begin{split}
		P_0\pwket=\omega\pwket,& \ P_0\pwketbar=\bar{\omega}\pwketbar,\\
		P_i\pwket=p_i\pwket,& \ P_i\pwketbar=\bar{p}_i\pwketbar,\\
		\omega^2=\sum_{i=1}^{3} p_i^2,& \ \bar{\omega}^2=\sum_{i=1}^{3} \bar{p}_i^2,
	\end{split}
\end{equation}
we find that the result of (\ref{eq:m}) is:
\begin{widetext}
\begin{equation}
	\label{eq:m2}
		\begin{split}
			-\left(\omega^2+\bar{\omega}^2+2\omega\bar{\omega}-\sum_{i=1}^3\left(p_i^2+\bar{p}^2_i+2p_i\bar{p}_i\right)\right)&\left(\pwket\otimes\pwketbar+\pwketbar\otimes\pwket\right)=\\
		-2\left(\omega\bar{\omega}-\sum_{i=1}^3p_i\bar{p}_i\right) &\left(\pwket\otimes\pwketbar+\pwketbar\otimes\pwket\right).
		\end{split}
\end{equation}
\end{widetext}
Since $\omega=|\pp|$ and $\bar{\omega}=|\ppbar|$, the term between brackets in the last line of Eq. (\ref{eq:m2}) can be written as $|\pp||\ppbar|-\pp\cdot\ppbar$. To get a mass squared eigenvalue equal to zero we then need 
\begin{equation}
	|\pp||\ppbar|=\pp\cdot\ppbar,
\end{equation}
which is only met if $\ppbar=\alpha\pp$ for real $\alpha>0$. Therefore, for a state of the kind (\ref{eq:boson}) to have zero mass, the two momenta $\pp$ and $\ppbar$ must be parallel.

Let us now look at the polarization of this kind of states using the helicity basis. Given a $\pp$ and a $\ppbar$ that fulfill the zero mass condition, the possible states of the kind (\ref{eq:boson}) are:
\begin{equation}
	\label{eq:three}
	\begin{split}
		&|\pp \ +\rangle\otimes|\alpha\pp \ +\rangle+ |\alpha\pp \ +\rangle\otimes|\pp \ +\rangle,\\
		&|\pp \ -\rangle\otimes|\alpha\pp \ -\rangle+ |\alpha\pp \ -\rangle\otimes|\pp \ -\rangle,\\
		&|\pp \ +\rangle\otimes|\alpha\pp \ -\rangle+ |\alpha\pp \ -\rangle\otimes|\pp \ +\rangle,\\
		&|\pp \ -\rangle\otimes|\alpha\pp \ +\rangle+ |\alpha\pp \ +\rangle\otimes|\pp \ -\rangle.\\
	\end{split}
\end{equation}
The application of the helicity operator $\prescript{}{2}{\Lambda}=\Lambda\otimes I + I \otimes \Lambda$ to these states shows that the four of them are eigenstates of helicity with eigenvalues $+$2, $-$2, 0 and 0, respectively. The last two do not correspond to a gravitational wave and must be discarded for our purposes. 

Applying the $\prescript{}{2}{P}_\mu$ operators shows that the states in Eq. (\ref{eq:three}) have four-momentum eigenvalues equal to $(1+\alpha)p_\mu$. At this point we can affirm that, given an arbitrary gravitational plane wave of helicity $\lambda^g$ and four-momentum $q_\mu$, the following set of electromagnetic tensor waves is equivalent to the gravitational wave:
\begin{equation}
\label{eq:ETW}
	|\kappa \mathbf{q} \ \frac{\lambda^g}{2}\rangle\otimes|(1-\kappa) \mathbf{q} \ \frac{\lambda^g}{2}\rangle
	+
	|(1-\kappa) \mathbf{q} \ \frac{\lambda^g}{2}\rangle\otimes|\kappa\mathbf{q} \ \frac{\lambda^g}{2}\rangle,
\end{equation}
for any real $\kappa\in(0,1)$. The allowed values of $\kappa$ follow from the massless condition $\kappa\mathbf{q}=\alpha(1-\kappa)\mathbf{q}$, with real $\alpha>0$.

\begin{figure*}[ht!]
	\begin{center}
	\includegraphics[width=0.7\linewidth]{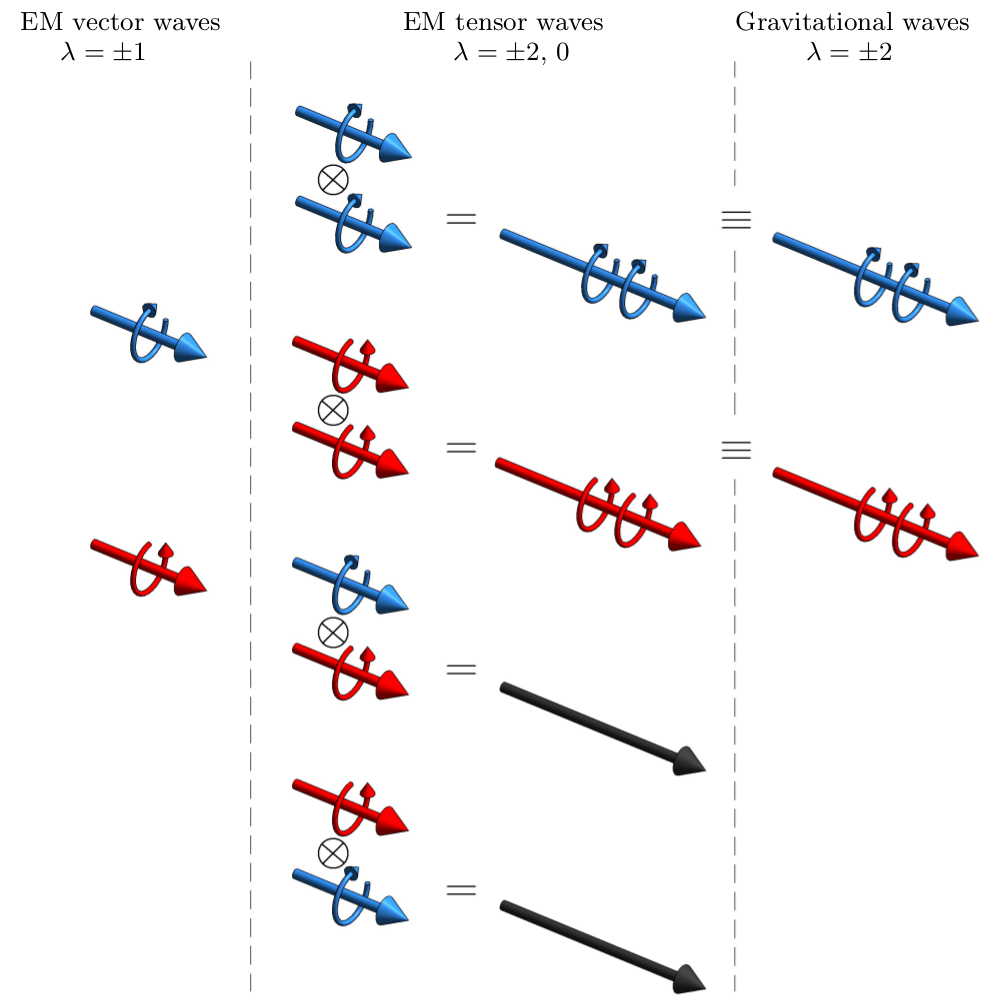}
	\caption{Equivalence between electromagnetic tensor product states and gravitational waves. The different tensor products of electromagnetic plane waves of well-defined helicity equal to $\pm1$ (in the left panel) result in electromagnetic tensor waves of helicity $\pm2$ and $0$ (middle panel), see Eq. (\ref{eq:three}). Only the first two have a gravitational wave equivalent (right panel). The momenta of the two electromagnetic plane waves do not need to be equal: Parallel momenta is enough to ensure that the tensor product will have a gravitational wave equivalent.}
\label{fig:em_grav_eq}
\end{center}
\end{figure*}

Figure \ref{fig:em_grav_eq} illustrates the results of the derivation.

These results are consistent with and get reinforced by group theoretical arguments involving the decomposition of tensor products of two identical massless representations of the Poincare group. The reduction of the direct product of two massless representations with the same helicity $\lambda$ results in both massless and massive irreducible representations \cite{Lomont1960}. All of the massless representations have helicity equal to $2\lambda$, and their multiplicity is infinite and uncountable, following the range of a real and positive constant \cite[Eq. 7.1]{Lomont1960}, which corresponds exactly to our parameter $\alpha$.

Since any gravitational wave solution of Eq. (\ref{eq:delta}) in flat spacetime can be written as a linear combination of gravitational plane waves of well-defined helicity, we conclude that any gravitational wave in a flat background is equivalent to a sum of electromagnetic tensor waves. Each electromagnetic tensor wave is the tensor product of two plane waves of the same helicity and parallel momentum.  

Equations (\ref{eq:splita}) and (\ref{eq:splitb}) can now be interpreted in light of these results. By re-writing the right hand side of Eq. (\ref{eq:splita}) in an exchange symmetric form (similarly for Eq. (\ref{eq:splitb}))

{\small
\begin{equation}
	\label{eq:auhome}
	\begin{split}
		\frac{1}{2}(\xhat &+ i\yhat)\exp(i\kappa^+\omega(z-t))\otimes(\xhat+i\yhat)\exp(i(1-\kappa^+)\omega(z-t))\\  
		+\frac{1}{2}(\xhat &+ i\yhat)\exp(i(1-\kappa^+)\omega(z-t))\otimes(\xhat+i\yhat)\exp(i\kappa^+\omega(z-t)),
	\end{split}
\end{equation}
}
we obtain a representation of the state 
\begin{equation}
|\kappa^+\omega\zhat\ +\rangle|(1-\kappa^+)\omega\zhat +\rangle+|(1-\kappa^+)\omega\zhat\ +\rangle|\kappa^+\omega\zhat +\rangle,
\end{equation}
restricted to equal position and time coordinates ($z$ and $t$) in both factors of the tensor products. This restriction is needed to fully equate measurements on two-body objects to measurements on single-body objects.

\section*{Generalization to curved spacetime backgrounds}\label{sec:curved}

We will now show that these flat spacetime results also hold in the case of gravitational waves propagating in a non-flat background, provided that the waves have a small amplitude and vary rapidly with respect to the background metric $\bmunu$, as indicated by Eq. (\ref{eq:regime}). With these restrictions, we go back to the approximate solution of Eq. (\ref{eq:ansatz}), $\hmunu(x_\alpha)=B_{\mu\nu}(x_\alpha)\exp(i\phi(x_\alpha))$, and start by exploiting the conditions that the polarizability tensor must meet. In the chosen gauge, $B_{\mu\nu}(x_\alpha)$ is transverse to the phase gradient four-vector $p_\mu(x_\alpha)=\partial_\mu \phi(x_\alpha)$, and purely spatial. This leaves a space which can be expanded using tensor products of two linearly independent spatial vectors. Since the tensor must also be symmetric, we can write:
\begin{equation}
	\label{eq:aiai0}
	\begin{split}
		B_{\mu\nu}(x_\alpha)&=c_1(x_\alpha)\left[\ehat_1(x_\alpha)\otimes\ehat_1(x_\alpha)\right]\\&+c_2(x_\alpha)\left[\ehat_2(x_\alpha)\otimes\ehat_2(x_\alpha)\right]\\&+c_3(x_\alpha)\left[\ehat_1(x_\alpha)\otimes\ehat_2(x_\alpha)+\ehat_2(x_\alpha)\otimes\ehat_1(x_\alpha)\right],
	\end{split}
\end{equation}
where $c_i(x_\alpha)$ are complex scalars. We also know that gravitational waves have only two gauge invariant degrees of freedom, and that the transverse traceless gauge isolates them. Therefore, expression (\ref{eq:aiai0}) has one degree of freedom too many. In order to eliminate it, we will impose that the two degrees of freedom correspond to helicity eigenstates with eigenvalues $\pm2$. This demand is consistent with the fact that the equations of general relativity in vacuum are invariant under the gravitational version of the duality transformations \cite[Sec. 3]{Maartens1998}. It follows that their solutions can be classified by the eigenvalues of the generator of duality transformations: Helicity \cite{Calkin1965,Zwanziger1968}. We hence require the existence of waves $\hmunu^{\pm}(x_\alpha)$ that are invariant under rotations by an arbitrary angle $\theta$ along the spatial momentum direction ($R_\pp(\theta)$), and pick up a factor of $\exp(\mp i2\theta)$. In the geometric optics approximation, the spatial momentum direction is given by the spatial part of $p_\mu(x_\alpha)$. The requirement of definite helicity equal to $\pm$2 reads then:
\begin{equation}
R_{\hat{\pp}(x_\alpha)}(\theta)\hmunu^{\pm}(x_\alpha)=\exp(\mp i2\theta)\hmunu^{\pm}(x_\alpha).
\end{equation}
Our requirement is met by assigning the first two degrees of freedom in Eq. (\ref{eq:aiai0}) to the two helicities $(\pm2)$ and discarding the third one. The required transformation properties of the tensors are then:
\begin{equation}
	\begin{split}
		&R_{\hat{\pp}(x_\alpha)}(\theta)\left[\ehat_{\pm}(x_\alpha)\otimes\ehat_{\pm}(x_\alpha)\right]=\\ 
		&\left[R_{\hat{\pp}(x_\alpha)}(\theta)\ehat_{\pm}(x_\alpha)\right]\otimes\left[R_{\hat{\pp}(x_\alpha)}(\theta)\ehat_{\pm}(x_\alpha)\right]=\\
		&\exp(\mp i2\theta)\left[\ehat_{\pm}(x_\alpha)\otimes\ehat_{\pm}(x_\alpha)\right], 
    \end{split}
\end{equation}
which means that the two vectors $\ehat_{\pm}(x_\alpha)$ must meet
\begin{equation}
	\label{eq:a}
	R_{\hat{\pp}(x_\alpha)}(\theta)\ehat_{\pm}(x_\alpha)=\exp(\mp i\theta)\ehat_{\pm}(x_\alpha).
\end{equation}
With this choice of $\ehat_{1}=\ehat_+$ and $\ehat_{2}=\ehat_-$, the discarded degree of freedom is, as in the flat background case, a helicity eigenstate with zero eigenvalue. 

With these results and the insight from the flat spacetime case we can write:
\begin{widetext}
\begin{equation}
	\label{eq:nonflat}
	\begin{split}
			\hmunu(x_\alpha)&=B_{\mu\nu}(x_\alpha)\exp(i\phi(x_\alpha))=\\
						 &\frac{c_{+}(x_\alpha)}{2}\left[\ehat_{+}(x_\alpha)\exp(i(1-\kappa^+)\phi(x_\alpha))\otimes\ehat_{+}(x_\alpha)\exp(i\kappa^+\phi(x_\alpha))\right]+\\
		&\frac{c_{+}(x_\alpha)}{2}\left[\ehat_{+}(x_\alpha)\exp(i\kappa^+\phi(x_\alpha))\otimes\ehat_{+}(x_\alpha)\exp(i(1-\kappa^+)\phi(x_\alpha))\right]+\\
		&\frac{c_{-}(x_\alpha)}{2}\left[\ehat_{-}(x_\alpha)\exp(i(1-\kappa^-)\phi(x_\alpha))\otimes\ehat_{-}(x_\alpha)\exp(i\kappa^-\phi(x_\alpha))\right]+\\
		&\frac{c_{-}(x_\alpha)}{2}\left[\ehat_{-}(x_\alpha)\exp(i\kappa^-\phi(x_\alpha))\otimes\ehat_{-}(x_\alpha)\exp(i(1-\kappa^-)\phi(x_\alpha))\right] \! = \! \hmunu^+(x_\alpha)+\hmunu^-(x_\alpha).
	\end{split}
\end{equation}
\end{widetext}
According to Eq. (\ref{eq:a}) the constituent vector waves in Eq. (\ref{eq:nonflat}) are helicity eigenstates with eigenvalues $\pm$1. According to Eq. (\ref{eq:nonflat}), their four-momenta have zero length because they are proportional to $p_\mu(x_\alpha)$. We recognize them again as electromagnetic waves, this time in the geometric optics approximation. We note that the decomposition of the polarization tensor into two components of definite helicity can be also achieved by taking the gauge invariant part of the null vierbein decomposition of $B_{\mu\nu}$ that appears in \cite[Eq. 5.17]{Ramos2006}. 

Let us now consider the propagation of the waves in Eq. (\ref{eq:nonflat}). We recall that in the geometric optics approximation, propagation is accomplished by parallel transporting both the phase gradient four-vector $p_\mu$ and the polarization tensor along the null geodesics $dx^\mu/dl=p^{\mu}$. All the tensors in Eq. (\ref{eq:nonflat}) have the same total phase attached to them. They will be transported along the same path. This null geodesic ray must also be the path of parallel transport for the constitutive vector waves, otherwise the factors would split up and break the required structure. This imposes $\kappa^{\pm}\in(0,1)$, since if either $\kappa$ or $(1-\kappa)$ were negative, the corresponding factor would propagate backwards along the geodesic. In the case of flat spacetime, we obtained the same allowed range of values for $\kappa^{\pm}$ from requiring massless states. The requisite of a consistent null geodesic ray is the massless condition in geometric optics. When higher order terms are added to the geometric optics solution, the two helicity components do not follow the same path \cite[II.3.5]{Iorio2007}. Nevertheless, since the vectorial factors in each helicity should still propagate together, the same restriction $\kappa^{\pm}\in(0,1)$ is needed.

The exact value of $\kappa^{\pm}$ does not change the null ray or the helicity. Its meaning is hence the same as in the flat spacetime case: It collects the set of electromagnetic tensor waves which are equivalent to a single gravitational wave.

Let us now investigate the effects of propagation in the factorized polarization tensor. We write the parallel transport equation for the total polarization tensor in (\ref{eq:nonflat}) by means of the covariant derivatives $D_{\mu}$: 
\begin{equation}
	\label{eq:pt}
	0=p^\sigma D_{\sigma}\left[c_+\ehat_+\otimes\ehat_+ + c_-\ehat_-\otimes\ehat_-\right].
\end{equation}
The first order linear differential equation (\ref{eq:pt}) with the initial condition at some initial point $x_\alpha=x_\beta$ has a unique solution at each point of the path. We now make the guess that the solution can be obtained by first parallel transporting both $\ehat_{\pm}$ and then taking their tensor products. This guess turns out to be correct. To verify it, we expand the covariant derivatives in (\ref{eq:pt}) across the tensor products:
\begin{equation}
	\label{eq:pt2}
	\begin{split}
	0=&c_+(x_\beta)\left(\left[p^\sigma D_{\sigma}\left(\ehat_{+}\right)\right]\otimes \ehat_{+}+\ehat_{+}\otimes\left[p^\sigma D_{\sigma}\left(\ehat_{+}\right)\right]\right)+\\
	&c_-(x_\beta)\left(\left[p^\sigma D_{\sigma}\left(\ehat_{-}\right)\right]\otimes \ehat_{-}+\ehat_{-}\otimes\left[p^\sigma D_{\sigma}\left(\ehat_{-}\right)\right]\right).
	\end{split}
\end{equation}
Note that the covariant derivatives do not act on the scalars $c_{\pm}(x_\beta)$, which are fixed initial conditions. Our guess assumes that the parallel transport equations for the polarization vectors $\ehat_{\pm}$ are met, i.e:
\begin{equation}
	\label{eq:pt3}
p^\sigma D_{\sigma}\left(\ehat_{\pm}\right)=0,
\end{equation}
which is clearly a solution of Eq. (\ref{eq:pt2}).

We can hence conclude that, in order to propagate the tensorial gravitational wave in Eq. (\ref{eq:nonflat}), we may as well take the constitutive electromagnetic waves in the tensor product factors, propagate them in the curved background by the rules of geometric optics, take the tensor products of the results and add both tensor products weighted by $c_{\pm}(x_\beta)$. The two helicity components of an electromagnetic vector wave do not mix during parallel transport. They pick up phases with the same absolute value and different sign \cite{Brodutch2011,Brodutch2011b}. The same is true for gravitational waves. The factorization allows us to recover the known result that the phase for the gravitational case is exactly twice that of the electromagnetic wave \cite[Sec. IV]{Ramos2006}.

All the above arguments are valid for any initial $x_\beta$ and $p_\mu(x_\beta)$, so they apply in all spacetime to each of the terms in a sum of an arbitrary number of solutions of the type $B_{\mu\nu}\exp(i\phi)$:
\begin{equation}
	\label{eq:general}
	\hmunu=\sum_n B_{\mu\nu}(n)\exp(i\phi(n)).
\end{equation}

This finishes the proof of the equivalence: Any gravitational wave of the kind (\ref{eq:general}) propagating in curved spacetime is equivalent to a sum of electromagnetic tensor product waves propagating in the same spacetime. Each term of the sum is the tensor product of two electromagnetic waves of the same helicity and parallel momentum. These results hold under the assumptions of small amplitude and fast variation of the gravitational wave with respect to the background metric (see Eq. (\ref{eq:regime}). 

We finalize this discussion by noting that the expression of the stress-energy tensor of the waves does not depend on whether they are gravitational or electromagnetic (see \cite[Eq. 35.77j]{Misner1973}, \cite[\S 4,\S 6]{Isaacson1968b}). Therefore, the portion of the background curvature due to the waves themselves is also equivalent in both cases, and extends the equivalence to the back action of the waves onto the background.

Throughout the analysis, we have assumed propagation in empty space. The different couplings of the electromagnetic and gravitational fields to material particles, for example to electrically charged particles, indicates that the equivalence will not hold in the presence of matter.

\section*{Quantum emulation of gravitational waves with entangled photon states}

We now use these results to propose a path for the quantum emulation of the propagation of gravitational waves in a laboratory by means of experimental electromagnetic setups. Besides the equivalence that we have shown, we will need two more ingredients: The conformal invariance of electromagnetism, and the fact that Maxwell's equations in an empty but curved spacetime are identical to Maxwell's equations in a material medium in flat spacetime.

\begin{figure*}[ht!]
	\begin{center}
	\includegraphics[width=0.7\linewidth]{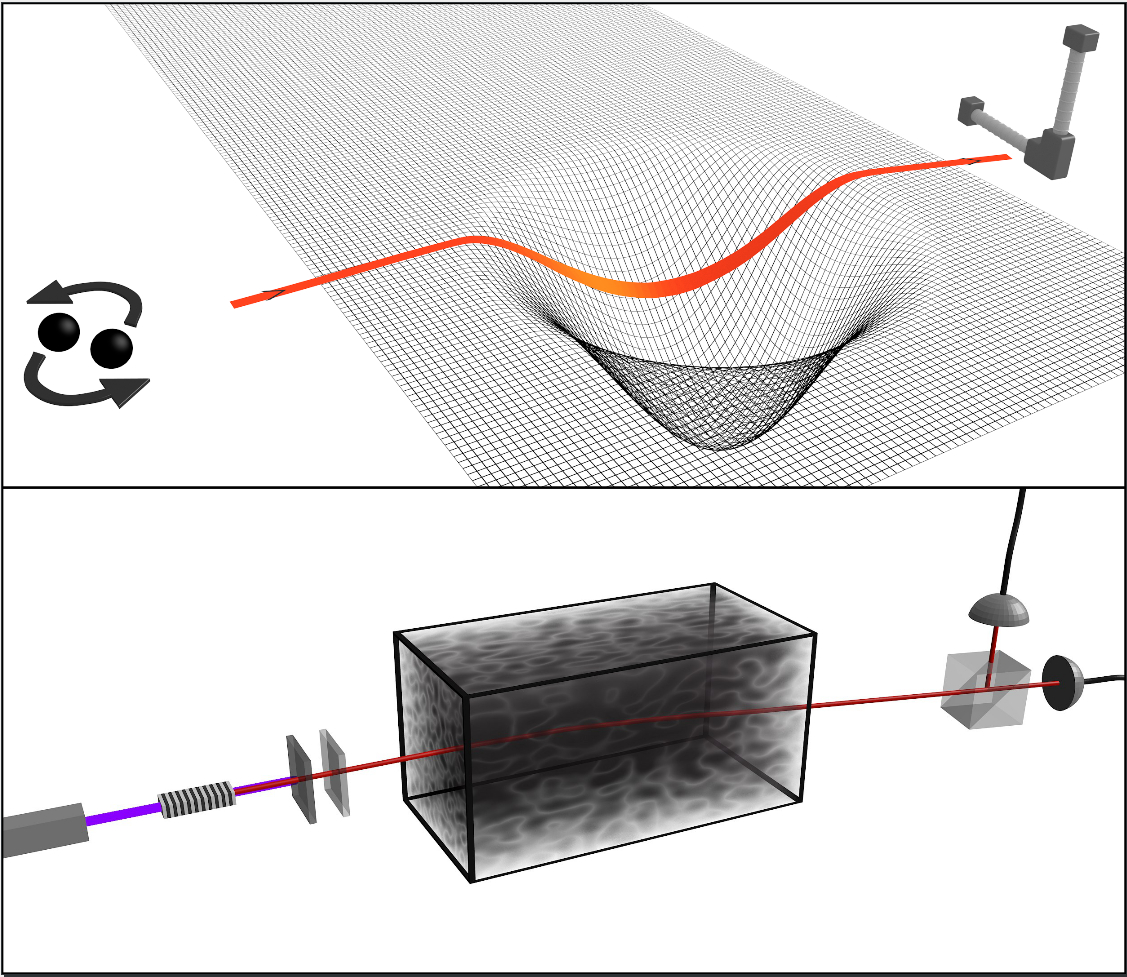}
\caption{Schematic comparison of a gravitational wave propagating through curved spacetime (upper panel) with the equivalent situation of two-photon states (lower panel) propagating through a metamaterial. The propagation of the gravitational wave is equivalent to the propagation of a subset of two-photon states. Conformal invariance is used to shrink the astronomical sizes down to laboratory scales, and a metamaterial media to mimic the specific spacetime curvature. The source for the two-photon states is a non-linear crystal, for example a periodically poled potassium titanyl phosphate (ppKTP), pumped by a narrow band laser. An interference filter after the crystal removes the pump light and could restrict the frequencies for the down-converted photons. A quarter waveplate would finally transform the state to the helicity basis. In order to emulate an arbitrary gravitational wave polarization, superpositions of photon pairs with both helicities are required. This can be achieved by combining two sources with orthogonal linear polarizations before the quarter waveplate. }
\label{fig:space_lab_comp}
\end{center}
\end{figure*}

Let us assume that we are interested in emulating a given scenario where gravitational waves from one or more sources propagate through curved spacetime regions (see the upper panel of Fig. \ref{fig:space_lab_comp}). The wavelengths of the detectable gravitational field are typically very large, larger than 10$^5$ meters \cite{Thorne1997}, as are the features of the curved background. Using the results contained in this article, we can map the scenario of interest to one where the gravitational waves are substituted by electromagnetic tensor waves. Now, the conformal invariance will allow us to shrink the problem. One of the symmetries of conformal invariance is a pointwise scaling of spacetime: $\bar{x}_\alpha=s(x_\alpha) x_\alpha$. We will use a constant $s(x_\alpha)=s$. Let us say that we want to study the solutions of Maxwell's equations in a spacetime with metric $\bmunu(x_\alpha)$. The invariance under scalings means that we can equivalently study the solutions of the system obtained after the change $\bar{x}_\alpha=s x_\alpha$, whose metric is $s^{-2}\bmunu(\bar{x}_\alpha/s)$. In terms of the electromagnetic field tensors, the solutions of the original and scaled system are related as $\bar{F}_{\mu\nu}(\bar{x}_\alpha)=s^{-2}F_{\mu\nu}(\bar{x}_\alpha/s)$. Features of size $L$ in the original spacetime geometry have size $s L$ in the scaled system, and the same relationship holds for the wavelengths in the two systems. After shrinking the problem to a suitable size, we are left with the task of creating a spacetime curvature $s^{-2}\bmunu$ in the laboratory. Fortunately, we do not need to tackle this problem directly: The propagation of electromagnetic waves in a given empty but curved spacetime is equivalent to their propagation in a particular material medium in flat spacetime \cite{Plebanski1960}. The constitutive relations of the material are a function of the metric and, in general, describe a continuous, inhomogeneous and anisotropic medium. Provided that one can fabricate such media, it is possible to emulate the propagation and interference of an arbitrary number of gravitational waves traveling through the spacetime geometry created by an arbitrary number of cosmological objects by means of experimental electromagnetic setups. Figure \ref{fig:space_lab_comp} illustrates this quantum emulation. It is important to note that the spacetime curvature caused by the gravitational waves themselves is not dynamically produced in this emulation framework. That is, the part of $\bmunu$ attributable to them must be small enough so that it can be ignored.

A technical challenge for this quantum emulation framework is the fabrication of the appropriate inhomogeneous and anisotropic media that mimic the desired spacetime geometry. Their design and fabrication is actively being investigated by the metamaterials community. These media implement transformation devices like for example cloaks and perfect lenses \cite{Leonhardt2006,Wegener2010,Soukoulis2011}. Actually, the helicity preservation inherent to a general curved spacetime \cite[\S 11]{Birula1996} can be exploited to provide design guidelines for transformation devices \cite{FerCor2013}. 

Having solved the scale problem and the propagation in a curved spacetime, we now turn to the generation of electromagnetic fields with the desired properties: That they can be decomposed as sums of tensor products of two electromagnetic plane waves of the same helicity, whose momentum four-vectors are parallel and proportional to one another. A system which already fulfills most of these properties is a spontaneous parametric down-conversion (SPDC) source of momentum correlated photons. Entangled photon sources based on SPDC generated photons have been regularly used in quantum optics experiments in the last 40 years. In these quantum light sources, a quadratic nonlinear crystal is pumped by a high frequency beam of light, typically in the ultraviolet or violet regime. The nonlinear crystal then mediates the low efficiency process of down-converting  some of the pump photons into two photons at a lower frequency (typically in the infra-red). The possibility of engineering the phase matching of nonlinear crystals has allowed for a wide range of engineered quantum states of light.

In particular, the use of collinear sources, i.e. sources where the generated photons are emitted along the propagation direction of the pump beam, allows for a tight control of the momentum and polarization correlations of the down-converted photons. Collinear sources based on Type 0 periodically poled potassium titanyl phosphate (ppKTP) can generate entangled photons which are momentum correlated and have the same linear polarization. Their polarization can then be transformed to the required helicity basis with a quarter waveplate. The momentum correlation is achieved by using a pump beam with a width which is much smaller than $\sqrt{S/k_p}$, where $S$ is the length of the crystal and $k_p$ the wavenumber of the pump beam. In the two-photon state, the terms exhibiting correlations between non-parallel momenta cannot be completely suppressed, but can be made arbitrarily small by either increasing the length of the crystal or using a tighter pump focus \cite{Leach2012}.

In order to emulate an arbitrary gravitational wave, superpositions of photon pairs with both helicities are required. This can be achieved by combining two of the described sources with orthogonal linear polarizations before the quarter waveplate. 

We note that the parameter $\kappa\in(0,1)$ in Eq. (\ref{eq:ETW}), which ties together all the electromagnetic tensor states that are equivalent to a single gravitational wave, has a physical correspondence in the source of momentum correlated photons. It corresponds to the split of the frequency of the pump $w_p$, which we assume monochromatic, into the frequencies of the emitted photons ($\omega_s,\omega_i$). Since energy conservation enforces $\omega_p=\omega_s+\omega_i$, we may write $\omega_s=\kappa\omega_p$ and $\omega_i=(1-\kappa)\omega_p$ for $\kappa\in(0,1)$, which makes the correspondence clear. Should the metamaterial exhibit a significant frequency dependence, $\kappa^{\pm}=1/2$ can be chosen by means of a narrow-band interference filter selecting the frequency degenerate down-converted photon. This completes the design of a quantum source of light, capable of emulating a gravitational wave.

For the detection of the field after it has passed through the metamaterial, we need to keep in mind that we are emulating single-body objects by means of two-body objects. The measurements should rely on detecting both photons ``at the same place and at the same time''. This may be done by counting coincidences, where both photons of a pair arrive at the detector within a very small time interval. Measurements that are unique to a two-particle state, like quantum interference, have no correspondence in the gravitational wave case, which places them outside the emulation framework.

The applicability of the emulation scheme to a particular background metric is determined by the applicability of the theoretical results contained in the previous section: The amplitudes of the gravitational waves must be much smaller than those of the background, and, the waves must vary much faster than the background (see Eq. (\ref{eq:regime})). For a given wavelength, this last condition limits the maximum allowable curvature. When the curvature increases, e.g. while approaching a singularity, the applicability reduces to correspondingly smaller wavelengths. 

\section*{Discussion}
In conclusion, we have shown that the propagation of gravitational waves in a curved spacetime background is equivalent to the propagation of a restricted set of electromagnetic tensor product waves. The defining characteristics of gravitational waves, zero mass and possible helicities equal to $\pm 2$, select the appropriate subspace of electromagnetic states. They turn out to be the tensor products of two electromagnetic waves of the same helicity and parallel momentum. Therefore, any linearly or elliptically polarized gravitational wave can be expanded as the sum of two electromagnetic tensor waves, one for each circular polarization handedness. The consideration of helicity eigenstates of both the gravitational and the electromagnetic waves has been crucial: A linearly polarized gravitational wave cannot be formally factorized as a tensor product of two electromagnetic waves, while a circularly polarized gravitational wave can.  Our analysis is restricted to gravitational waves of small amplitude that vary rapidly with respect to the spacetime background and propagate in empty space. As far as we know, this equivalence has not been reported before.

We have shown that the implementation of the equivalence needs genuine quantum correlations between two photon states. As an application of these results, we have proposed a path for the quantum emulation of the propagation of gravitational waves by means of experimental electromagnetic setups. First, the equivalence derived in this article is applied to map the gravitational waves into electromagnetic tensor waves, then the conformal invariance of electromagnetism is used to shrink the problem from large wavelengths and large sized spacetime features down to suitable sizes, and finally, the scaled down curved spacetime background is emulated in the laboratory by an appropriate material medium. Precisely this type of media are already the object of research in the field of metamaterials.

To finish, we suggest that the decomposition of gravitational waves into tensor products of electromagnetic waves could be used to reduce the complexity of simulating gravitational waves in a computer. If the scenario of interest meets the specified restrictions, setting $\kappa^{\pm}=1/2$ in Eq. (\ref{eq:nonflat}), one can first simulate the propagation of two electromagnetic vector waves, one for each helicity, and then sum the tensor products of each resulting vector wave with itself. 

Future expansions of this framework could include its extension outside the geometrical optics approximation in order to study the interaction of gravitational waves with strongly curved spacetimes, and the use of nonlinear optical metamaterials capable of reproducing the back action of the gravitational wave on the background metric.

\section*{Acknowledgments}
This work was funded by the Center of Excellence for Engineered Quantum Systems and by the Basque Government IT472- 10 Grant, Spanish MINECO FIS2012-36673-C03-02, Ram\'on y Cajal Grant RYC-2012-11391, UPV/EHU UFI 11/55, UPV/EHU EHUA14/04, CCQED, PROMISCE and SCALEQIT European projects. G.M.-T is also funded by the Future Fellowship program (FF).
I.F.-C wishes to acknowledge Ms Magda Felo for drawing Fig. 1. 

\section*{Author contributions}
GMT and IFC contributed the initial idea of using entangled photon states to emulate gravitational waves. The initial idea was discussed and further developed together with ES and LL. IFC and MC derived the equivalence between electromagnetic tensor product states and gravitational waves in flat and curved spacetimes. AB contributed to the development of the experimental framework. All authors contributed to the writing of the manuscript.

\section*{Additional Information}
The authors declare no competing financial interests.


\end{document}